\documentclass[aps,prb,twocolumn,superscriptaddress,showpacs]{revtex4-1}
\pdfoutput=1

\usepackage{epstopdf}
\usepackage{epsfig}
\usepackage{amsmath}
\usepackage{amssymb}
\usepackage[colorlinks=true,citecolor=blue,linkcolor=blue]{hyperref}

\newcommand{\braket}[1]{\left<#1\right>}
\newcommand{\para}[1]{\left(#1\right)}

\newcommand{\ld}[0]{\ \ \ \ \ \ }
\newcommand{\sd}[0]{\ \ \ }

\begin{document}

\title{Electron-phonon interaction in the strongly correlated systems and its implications on the high Tc cuprates}

\author{Abolhassan Vaezi}
\email{vaezi@ipm.ir}
\affiliation{School of Physics, Institute for Research in Fundamental Sciences, IPM, Tehran, 19395-5531, Iran}

\begin{abstract}
The theory of electron-phonon interaction in the presence of strong correlation has been investigated in the present work. Due to the so called spin-charge separation, it is argued that the electron-phonon interaction in the strongly interacting systems is fundamentally different from the weakly interacting counterparts. Both spinons and holons are shown to carry electric charge and couple to the electromagnetic field and consequently to phonons.  A systematic approach to calculate the dynamical charge of spinons and holons is presented, followed by a discussion of implications of this model for the high temperature superconductors. It is demonstrated that the unusual oxygen isotope effect in high Tc materials can be easily understood within this framework.
\end{abstract}



\maketitle
\section{Introduction}
Providing a satisfactory description of the strongly correlating systems is one of the most challenging problems in theoretical condensed matter physics. The discovery of the high temperature superconductivity in cuprate compounds~\cite{Bednorz_Mueller_1986,Lee_Nagaosa_Wen_2006a} has made this challenge even more intriguing. Although, there has been a huge progress in this field during the past three decades, we still lack enough knowledge about the nature of the electron-phonon interaction in the strongly correlating systems. In the present paper, this issue is addressed from the spin-charge separation point of view and the effect of strong interaction on the coupling of electrons to phonons is investigated.

Although at the moment there is no general consensus about the underlying mechanism of superconductivity in cuprates and the glue that pairs up electrons, there are several theories that more or less capture the essence of physics in these materials. However, it should be mentioned that it is widely accepted that the physics of high Tc is that of doped Mott insulator. In this paper, we particularly work in the framework initially introduced by P. W. Anderson. According to the Anderson's idea of preformed Cooper pairs, the basic physics of high Tc cuprates is governed by strong correlation physics due to their proximity to the Mott insulator phase at half filling \cite{Anderson_1987Sci}. This approach is based on the resonating valence bond(RVB) state and the spin charge separation picture. At half filling, RVB state is made up of singlet Cooper pairs, but due to strong correlations, they do not have phase coherence and therefore, we obtain an insulating phase. Upon doping the system, they acquire the required phase coherence and we obtain a superconducting state.

This approach can explain basic properties of cuprates and yields a phase diagram similar to that of high Tc cuprates. On the other hand, observation of strong isotope effect on both the transition temperature and the superfluid density in cuprates \cite{Zimmerman_1,Khasanov_3_2008,Raffa_1,Zhao_1,Franck_1991,Keren_2011_a,Isotope}, indicates the importance of the electron-phonon interaction in understanding the physics of high Tc \cite{Schneider_Keller_1992_a,Batlogg_1,Phononic}. The reported oxygen isotope effect (OIE) is very different from that of conventional BCS superconductors in which the electron-phonon interaction is responsible for the pairing mechanism. These observations led us to reexamine the electron-phonon interaction in the presence of strong correlations and its effect on the pairing mechanism.

\begin{figure}[tbp]
\begin{center}
\includegraphics[width=180pt]{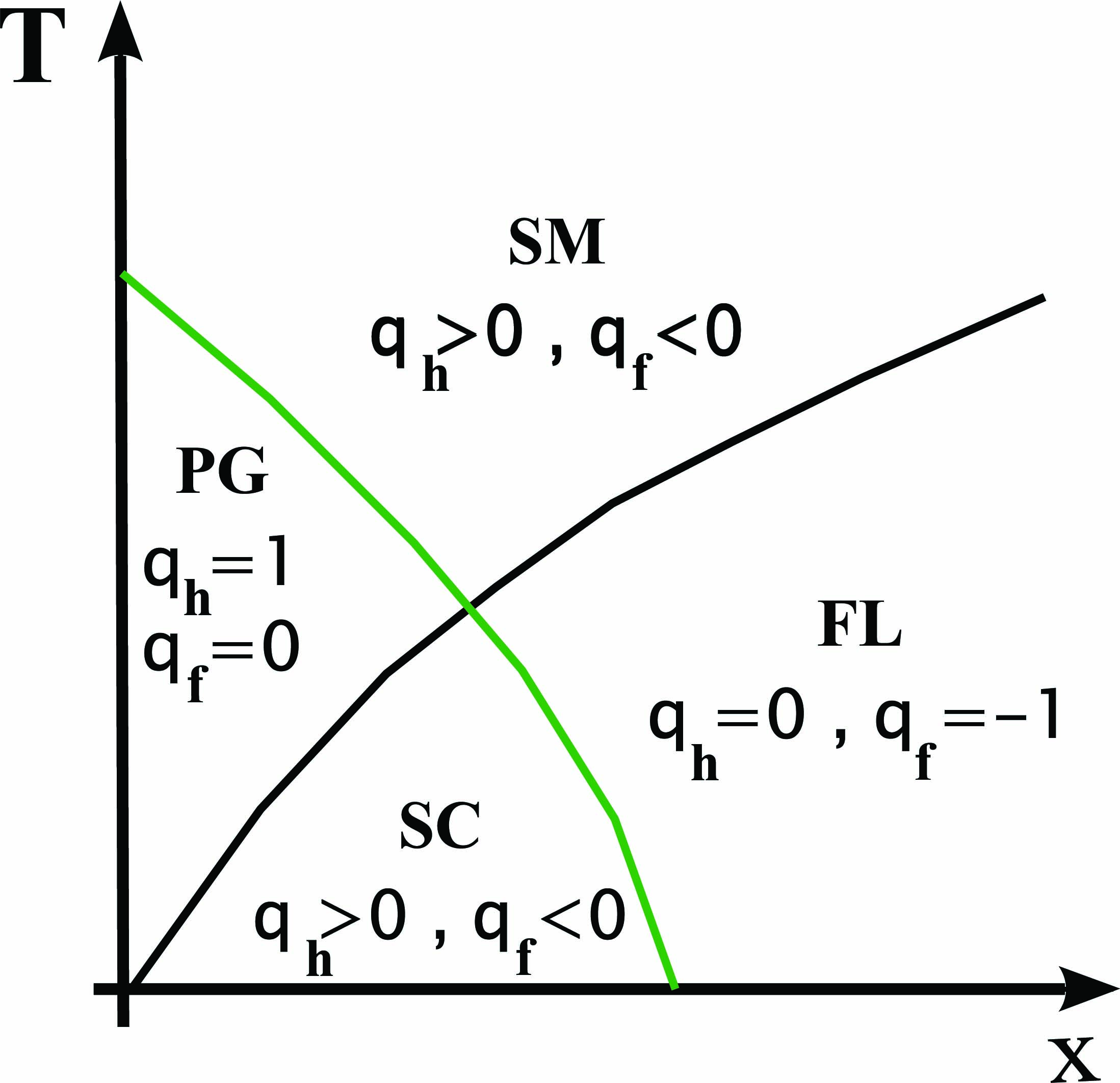}
\caption{(Color online) Schematic phase diagram and the electric charge of holons and spinons. SM stands for the strange metal phase where neither holons nor spinons condense. Sc is superconducting, FL fermi liquid and PG pseudogap phase. }\label{Fig}
\end{center}
\end{figure}

The deformation of a lattice increases its electric potential energy quadratically in the simplest approximation. Phonons are the quantum of these vibrations. A charged quasiparticle e.g. an electron couples to the phonon field proportional to its electric charge. Therefore, electrically neutral quasiparticles do not couple to phonons. A fundamental question about the electron-phonon interaction in the strongly correlated materials that was raised by the spin charge separation assumption is how to assign electric charge to spinons and holons. Do spinons carry all the electric charge and holons remain neutral? Should we assume the electric charge goes to holons and spinons are neutral quasiparticles? The main goal of this paper is to find the answer to this question unambiguously. The only assumption made here is the spin-charge separation and we demonstrate that the electric charge is a dynamical quantity that changes around the phase diagram of high Tc compounds. For example, in the Fermi liquid phase, we show that only spinons carry electric charge while in the pseudogap phase only holons carry the electric charge. In the superconducting region both holons and spinon carry an electric charge. In every region of the phase diagram $q_{h}-q_{s}=+1,$ where $q_{h}$ is the electric charge of holons and $q_{s}$ is that of spinons.

\section{Method}
In the slave boson method, electron operator at site $i$ with spin $\sigma$ is written in terms of slave particles as 

\begin{eqnarray}
  c_{i,\sigma}^\dag=f_{i,\sigma}^\dag h_{i}.
\end{eqnarray}  

Holons are assumed to be hard-core bosons. According, there is no doublon state in the Hilbert space and at each site we either have spinon or holon. In other words 

\begin{eqnarray}
n_{i,\uparrow}^{f}+n_{i,\downarrow}^{f}+n_{i}^{h}=1.  
\end{eqnarray}

The above definition for the electron operator is clearly invariant under local U(1) gauge transformation, provided spinons and holons carry the same charge with respect to the internal gauge field. By scaling the internal gauge field we can assume the internal charge of slave particles is ${\rm e_{int}= 1}$ (in units of electron charge ${\rm e}$). Due to the local constraint on the Hilbert space, hopping of a spinon from site $i$ to site $j$, should be accompanied by another hopping of a holons from site $j$ to site $i$. So we conclude that the current carried by spinons is equal but opposite to the current carried by holons and they add up to zero. Consequently, we have
\begin{eqnarray}
  {\bf J}^{f}_{\mu}\para{\vec{x}}+{\bf J}^{h}_{\mu}\para{\vec{x}}=0.
\end{eqnarray}

Now let us assign electric charges $q_{h}$ and $q_{f}$ to holons and spinons respectively. On the other hand the electron operator transforms under electromagnetic gauge transformation as $c_{i,\sigma}^\dag \to \exp\para{-i\alpha_i} c_{i,\sigma}^\dag$. But we have
\begin{eqnarray}
  c_{i,\sigma}^\dag =f_{i,\sigma}^\dag h_{i} \to &&\exp\para{i\para{q_{f}-q_{h}}\alpha_i}f_{i,\sigma}^\dag h_{i}\cr
  &&=\exp\para{i\para{q_{f}-q_{h}}\alpha_i}c_{i,\sigma}^\dag.
\end{eqnarray}

Therefore, we have
\begin{eqnarray}
  q_{h}-q_{f}=+1.
\end{eqnarray}

The above constraint does not specify the values for $q_f$ and $q_h$ independently and states that any value for $q_h$ and $q_f$ are accepted as long as they satisfy the above equation. In order to satisfy the constraint in Eq. [3], the external electromagnetic field has to generate an internal magnetic field proportional to itself as well. It is this internal magnetic field that finally resolves the dilemma related to the values of $q_h$ and $q_f$ and uniquely dictates their values. In the following we comment more on the procedure leading to this conclusion.

Let us assume a local electromagnetic field $A_{\mu}\para{x}$ and internal gauge field $a_{\mu}\para{x}$. Holons couple to $\para{q_{h}a_{\mu}\para{x}+q_{\rm int,h}a_{\mu}\para{x}}$ and spinons feel $\para{q_{f}A_{\mu}\para{x}+q_{\rm int,f}a_{\mu}\para{x}}$, where $q_{\rm int,h}=q_{\rm int,f}=1$. Within linear response theory, they generate current as
\begin{eqnarray}
  &&{\bf J}^{h}_{\mu}\para{x}=\Pi^{h}\para{q_{h}A_{\mu}\para{x}+a_{\mu}\para{x}},\cr
  &&{\bf J}^{f}_{\mu}\para{x}=\Pi^{s}\para{q_{f}A_{\mu}\para{x}+a_{\mu}\para{x}},
\end{eqnarray}
where we have assumed $-i\left[J_{f,h}^{\mu},J_{f,h}^{\nu}\right]=\Pi_{f,h} \delta_{\mu,\nu}$. Using equation [3], we have
\begin{eqnarray}
  a_{\mu}\para{x}=-\frac{q_{h}\Pi^{h}+q_{f}\Pi^{f}}{\Pi^{h}+\Pi^{f}}A_{\mu}\para{x}.
\end{eqnarray}

Substituting the above relation in Eq. [6],
\begin{eqnarray}
  &&{\bf J}^{h}_{\mu}\para{x}=\para{q_{h}-q_{f}}\frac{\Pi^{f}}{\Pi^{h}+\Pi^{f}}\Pi^{h}A_{\mu}\para{x},\cr
  &&{\bf J}^{f}_{\mu}\para{x}=-\para{q_{h}-q_{f}}\frac{\Pi^{h}}{\Pi^{h}+\Pi^{f}}\Pi^{f}A_{\mu}\para{x}
\end{eqnarray}

Now using the fact $q_{h}-q_{f}=1$, the above relations mean the physical response of the slave particles to the electromagnetic field is independent of the initial assumption for $q_h$ and $q_f$. However, we can unambiguously assign the following physical charges to spinons and holons due to their response to the electromagnetic field as ${\bf J}^{f,h}=q_{f,h}^{\rm eff}\Pi_{f,h} {\bf A}$
\begin{eqnarray}
  &&q^{\rm eff}_{h}=\frac{\Pi^{f}}{\Pi^{h}+\Pi^{f}}\cr
  &&q^{\rm eff}_{f}=q^{\rm eff}_{h}-1=-\frac{\Pi^{h}}{\Pi^{h}+\Pi^{f}},
\end{eqnarray}
which are independent of the details of charge assignation. From now on we drop ${\rm eff}$ symbol and simply use $q_{h}$ and $q_{f}$. Since $\Pi_{f,h}=-i\omega \sigma_{f,h}$, we can alternatively write
\begin{eqnarray}
  &&q_{h}=\frac{\sigma^{f}}{\sigma^{h}+\sigma^{f}},\cr
  &&q_{f}=q^{\rm eff}_{h}-1=-\frac{\sigma^{h}}{\sigma^{h}+\sigma^{f}}.
\end{eqnarray}

For the physical electric current we have 

\begin{eqnarray}
  {\bf J}_{ph}=q_{h}{\bf J}_{h}+q_{f}{\bf J}_f=\para{q_h-q_f}{\bf J}_{h}=\frac{\sigma_h \sigma_f}{\sigma_h+\sigma_f} {\bf E}
\end{eqnarray}

The above relation is known as Ioffe-Larkin recombination formula. Now let us consider different regions of the phase diagram of compute the effective charge of holons and spinons (see Fig. 1).

\subsection{\bf Fermi Liquid Phase} In the Fermi liquid (FL) phase, holons are condensed but spinons are not. So we have $\braket{h}\neq 0$ and $\Delta_{f}=0$, where $\Delta_f=\braket{f_{i,\uparrow}f_{j,\downarrow}}$ is proportional to the pseudogap energy. In this case, $\Pi_{h}={\rm superfluid~~ density}=\frac{n_{c,h}\para{x,T}}{m^*_{h}} \neq 0$ ( $n_{h,c}$ is the condensation fraction and $m^*_h$ effective holon mass, $x$ doping and $T$ temperature), while $\Pi_f=-i\omega \sigma_{f} \to 0$, we have

\begin{eqnarray}
  q_h= -\frac{\Pi_f}{\Pi_f+\Pi_h}\to 0, \ld q_{f}\to -1
\end{eqnarray}

Therefore, in the Fermi liquid we have $q_{f}\simeq -1$ and therefore, they have nonzero overlap with physical electrons and carry the same charge. On the other hand $q_{h}\simeq 0$ and we can safely assume that holons are electrically neutral in the FL phase. This can also be directly seen from $c_{i,\sigma}^\dag \simeq \braket{h} f_{i,\sigma}^\dag$. One important result of this argument is that in the absence of the pseudogap ({\em i.e.} when $\Delta_s=0$), condensed holons do not couple to phonons, since phonons only couple to electrically charged quasiparticles. In other words, phonons create local electromagnetic field and this field induces another (internal) gauge field. Holons couple to the sum of these two fields. The induced internal field almost cancels out the electric field and therefore, holons do not response to the electromagnetic field created by phonons. Therefore, when in the FL phase where holons condense and pseudogap is absent, ({\em i.e.} at low enough temperatures and large enough doping) only spinons respond to the phononic electromagnetic field. This argument is valid close to the phase transition between the Superconducting (SC) state and the FL state, i.e. in the overdoped region. In the next section we discuss that the phenomenology of cuprates inspires us to assume the isotope effect is solely due to the holon-phonon interaction. This immediately explain why there is no isotope effect on Tc in the overdoped region where SC state undergoes transition into the FL state.

\subsection{\bf Superconducting Phase} In the superconducting (SC) phase , both holons and spinons are condensed. So we have $\Delta_{h}=\braket{hh}\neq 0$ and $\Delta_{s}\neq 0$. In this case, since $\Pi=\frac{n_c\para{T}}{m^*}$, both $q_h$ and $q_f$ are nonzero. On the hand, for the d-wave superconductors we have

\begin{eqnarray}
  n_{c,f}\para{x,T}=n_{c,f}\para{x,0}\para{1-\frac{T}{T_{f}\para{x}}} \propto \Delta_{f}\para{x,T}^{\alpha},~~
\end{eqnarray}
 
with some exponent $\alpha \sim 2$. At zero temperature, almost all spinons condense and we obtain $n_{c,f} \sim 1-x$. For holons we have 

\begin{eqnarray}
  n_{c,h}\para{x,T}=x\para{1-\para{\frac{T}{T_{h}\para{x}}}^3}
\end{eqnarray}

Computing the effective charge of spinons and holons and we obtain $q_{f}\propto -\Pi_{h} \propto -x$ and $q_{h}\propto \Pi_f \propto \para{1-x}$ respectively. Therefore, in the superconducting state, spinons only respond to the $x$ fraction of the electromagnetic field. Since holons are effectively charged quasiparticles in this state, (or since the internal gauge field does not cancel out the local electromagnetic field of phonons completely) they couple to phonons. The larger pseudogap value, the stronger interaction between holons and phonons. This is an intuitive way to justify why the isotope effect is directly dependent to some power of the pseudogap energy in the experimental observations. This means that gauge fluctuations, renormalizes the coupling constant of holon-phonon interaction from the brae value $\gamma$, to $\gamma^*=\Delta_s^{\alpha} \gamma$. Using this expression and since $\Delta_{s}$ decreases with doping, we can easily explain why the isotope effect is a decreasing function of doping. Moreover, in the overdoped region, $\Delta_{s}=0$ at the transition temperature and therefore, holons do not interact with phonons. Subsequently, we do not expect isotope effect on the $T_c$ in the interface between SC and FL phases. However at $T=0$, $\Delta_{s}\neq 0$ even in the overdoped region and the mass of holons enhances in an isotope dependent way and this explains the nonvanishing isotope effect on the superfluid density and the London penetration depth in this region.

\subsection{\bf Pseudogap phase} In the pseudogap phase (PG), spinons form Cooper pairs with d-wave symmetry and condense. This region is above the underdoped SC region. In the underdoped side of the SC phase, according to phase fluctuation scenario for the SC transition, superconductivity disappears above $T_{\rm BKT}\para{x}$, the temperature above which Cooper pairs lose their long range phase coherence and their phase fluctuates strongly, vortices proliferate and superconductivity disappears. We assume this temperature is of the order of BEC temperature for the holon gas, but a bit less i.e. $T_{\rm BKT} ~ ^{<}_{\sim}~ T_{\rm BEC}$. Therefore, we assume in the PG region holons are still condensed between $T_{\rm BEC}\para{x}$ and $T_{\rm BKT}\para{x}$. Above $T_{\rm BKT}$ we have $\braket{h}=0$. Accordingly
\begin{eqnarray}
  &&T_{\rm BKT}<T<T_{\rm BEC}:\sd q_h\para{x,T}= \frac{\Pi_f}{\Pi_f+\Pi_h}<1, \sd q_{f} < 0 \cr
  &&q_h\para{x,T}\propto n_{c,f}\para{T,x}\propto \para{1-\frac{T}{T_f\para{x}}}\propto \Delta_{f}^2\para{T,x} 
\end{eqnarray}

However, above the BEC temperature, since the superfluid density of holon gas vanishes i.e. $\lim _{\omega\to 0}\Pi\para{q=0,\omega} \to 0$, we have

\begin{eqnarray}
  T>T_{\rm BEC}:\sd q_h\para{x,T}= \frac{\Pi_f}{\Pi_f+\Pi_h}=1, \sd q_{h}\to 0.
\end{eqnarray}

Therefore, deep in the PG region, only holons carry electric charge and couple to phonons. Spinons carry electric charge only in a narrow region between $T_{\rm BKT}\para{x}$ and $T_{\rm BEC}\para{x}$.

\section{Oxygen isotope effect in cuprates} 

It has been experimentally verified that $^{16}$O/$^{18}$O isotope substitution affects the SC transition temperature as well as the superfluid density. Since oxygen is lighter than copper in cuprate compounds and displaces easier, oxygen atoms contribute more to the lattice vibrations and cause isotope effect. Experimental observations indicate a strong OIE on $T_c$ only in underdoped cuprates(see Fig. 1), while there is no considerable OIE in overdoped cuprates. On the other hand OIE on the superfluid density has been reported for both sides \cite{Hofer_2000_a,Khasanov_1_2004,Tallon_1,Khasanov_4_2003}. In both cases the isotope exponent decreases as we approach the optimal doping from the underdoped side. On the other hand, there is no isotope effect on the effective mass of the nodal quasiparticles \cite{Iwasawa_1}. It is worth comparing these behaviors with that of conventional BCS superconductors. In conventional superconductors, there is no isotope effect on the superfluid density as well as the effective mass of quasiparticles, and the isotope effect on $T_c$ is independent of doping with isotope exponent around 1/2 \cite{BCS}. We would like to mention that the BCS theory and its generalizations e.g. Migdal-Eliashberg theory are based on the adiabatic approximation. Adiabatic limit assumes the phonon energy is much less than the typical energy of electrons i.e. $E_{\rm F}$. Therefore, adiabatic approximation in the Einstein model means $\omega_{_E}\ll E_{\rm F}$, where $\omega_{_E}$ is energy of Einstein phonon mode (optical phonons).
 
In high Tc cuprates, the situation is different if we assume spin charge separation. The typical energy of spinons is of the order of electrons. Experimental data implies that the kinetic energy of electron (and as a result that of spinons) is around the exchange energy $J$ which is about $1500$ Kelvin. On the other hand, the typical energy of phonons is expected to be several hundred Kelvin. Therefore, $E_{\rm F} > \omega_{_E}$ and we are in the regime of adiabatic spinon-phonon interaction. Subsequently, the mass of spinons renormalizes as $m^*_{f}=m^{f}\para{1+\lambda}$, where $\lambda$ is the coupling constant of spinon-phonon interaction. In the simplest model it can be shown that $\lambda \propto \frac{1}{M\omega_{_{E}}^2}$, and since $\omega_{_{E}}\propto \frac{1}{\sqrt{M_{ion}}}$, $\lambda$ is isotope independent. Therefore, oxygen isotope substitution does not enhance the effective mass of spinons. This is consistent with the experiment where there is no isotope effect on Fermi velocity by Laser ARPES, while there is shift in kink energy$~$\cite{Iwasawa_1}.

For holons we must be more careful. Holons are bosonic objects and their typical energy is very small compared to their bandwidth as long as they condense. Therefore, for holons we have $E_{h}<<\omega_{_E}$ and we should use the results of the non-adiabatic limit of the holon-phonon interaction.

To explain the isotope effect in cuprates, note that in the slave boson approach, superfluid density is inversely proportional to the mass of holons, $\rho_{\rm SC}(x,T=0) =\frac{x}{m^*_h}$. Since the isotope substitution changes the superfluid density in the whole region and decreases with increasing doping (it finally vanishes close to the FL phase), we conclude that the mass of holons should change with isotope substitution as well. On the other hand, $T_c$ is proportional to the superfluid density in the underdoped region and therefore,  $T_c(x)\propto 1/m^*_h$ in the underdoped side. This immediately means we can explain OIE on Tc provided we show OIE on $m^*_h$. On the other hand, Tc in the overdoped region is given by the transition temperature of spinons above which pseudogap closes.  Since antiferromagnetic exchange dominates over the phonon mediated attraction in cuprates, oxygen isotope substitution does not affect Tc much in the overdoped region.

From the above discussion we conclude that an isotope dependent holon mass can easily explain the observed OIE in cuprates. In the first half of this paper, we discussed that phonon field couples to charged quasiparticles. Therefore, holons can in principle couple to phonon if they acquire nonzero charge. In the previous section, we demonstrated that $q_{h}\para{x,T}\propto \rho_{f}\para{x,T} \propto \para{1-\frac{T}{T_f\para{x}}}$. Therefore, we can write down the following effective model for holons

\begin{eqnarray}
H=&& -\sum_{<i,j>}t_{h} h_{i}^\dag h_{j}-\sum_{i}\mu_{h}h_{i}^\dag h_{i} +\sum_{q} \omega_{_{E}} a_{q}^\dag a_{q}\cr
&&+\sum_{q,n,\sigma} q_{h}\frac{\gamma_{0}}{\sqrt{2M_{_O}\omega_{_E}}} h_{n}^\dag h_{n} \para{a_{n}^\dag +a_{n}},
\end{eqnarray}
where $\gamma_{0}$ is the coupling constant of electron-phonon interaction and $a_i^\dag$ ($a_i$) is the creation (annihilation) operator of phonons at site $i$. Note that $q_h$ factor renormalizes bare $\gamma_0$ to $q_h\gamma_0$. The above model has been extensively studied. In the non-adiabatic regime, we can use the results of single polaron theory \cite{Alexanrov_1,Alexanrov_2}. Therefore, we can do the powerful Lang-Firsov transformation \cite{Hohenadler_1}, {\em i.e.} $H\rightarrow \tilde{H}=e^{-S}He^{S}$ where:  $S=\frac{1}{{\omega_{_E}}\sqrt{2M\omega_{_{E}}}}\sum_{q,n} q_{h}\gamma_{0} h_{n}^\dag h_{n} \para{a_{n}^\dag -a_{n}}$. Under this transformation, $h_{i}$ transforms as:  $e^{-S}h_{i}e^{S}=h_{i}\mbox{exp}\para{q_h\frac{\gamma_{0}\para{a_{i}^\dag -a_{i}}}{{\omega_{_E}}\sqrt{2M\omega_{_{E}}}} }$. If we replace the exponential factor by its average, we finally have:

\begin{eqnarray}
  \tilde{H} = &&-\sum_{\vec{k}}\para{2t^*_{h}\para{\cos{k_x}+\cos{k_y}}+\mu^*_{h}}h_{k}^\dag h_{k}\cr
  &&+ \sum_{q}  \omega_{_E} \para{d_{q}^\dag d_{q}+1/2}+...
\end{eqnarray}
where ellipses stand for interaction terms and  

\begin{eqnarray}
  &&t^*_{h}\para{x,T}= \exp\para{-g^2\para{x,T}}t_{h}.\\
  && g^2\para{x,T} = q_{h}^2\para{x,T}\frac{\lambda_{0}}{ \omega_{_E}}.\\
  &&\lambda_{0}=\frac{\gamma_{0}^2}{M\omega_{_E}^2},
\end{eqnarray}
in which $M_{_O}$ is the oxygen mass. It is clear that $\lambda_0$ is mass independent from its definition. Now if we expand the energy of holons around $\vec{k}=0$ we have

\begin{eqnarray}
&&  \epsilon_{h}\para{k}=\frac{k^2}{2m_{h}^*}-\mu^{*}_h.\\
&&  m^{*}_{h}\para{x,T}=m_h \exp\para{g^{2}\para{x,T}},
\end{eqnarray}
where $m^*_h=\frac{1}{2t_h}$. As we see from the above equation, scattering off the phonons enhances the mass of holons by an exponential factor. This factors depends on the isotope mass and as a result we expect the OIE on any physical quantity depending on the mass of holons. The OIE on $m^{*}_h$ is

\begin{eqnarray}
\frac{d\log m^*_h}{d\log M_{_O}}=\frac{1}{2}g^2\para{x,T}.
\end{eqnarray}

\subsection{Oxygen isotope effect on the superfluid density}

The oxygen isotope effect on the superfluid density is defined as:
\begin{eqnarray}
  \beta\para{x}=-\frac{d \log \rho_s\para{x,T=0}}{d\log M_{_{O}}}.
\end{eqnarray}

According to the Ioffe-Larkin formula$~$\cite{Ioffe_1} the physical superfluid density is related to the superfluid density of spinons and holons as: $\rho_{ph}^{-1}=\rho_{h}^{-1}+\rho_{s}^{-1}$. Since condensation fraction of holons and spinons at zero temperature are $x$ and $1-x$ respectively, and from $\rho=\frac{n_{c}}{m^*}$ we have: $\rho_{ph}^{-1}\para{0}=\frac{m^{*}_{h}}{x}+\frac{m^{*}_{f}}{1-x}$. For small values of $x$, we have $\rho_{ph}\para{0}\simeq\frac{x}{m^{*}_{h}}$. Therefore, we have:

\begin{eqnarray}
  \beta\para{x}=\frac{d \log m^*_h\para{x,0}}{d\log M_{_{O}}}=q^2_{h}\para{x,0}\frac{\lambda_{0}}{\omega_{_{E}}}
\end{eqnarray}

Since $q_h\para{x,T}\propto \Pi_f\para{x,0}=n_{c,f}\para{x,0}/m^*_f$ and the condensation fraction of spinons decreases with increasing doping until vanishes as we approach the FL phase, $\beta$ reflects the same behavior which is consistent with empirical data. 

\subsection{Oxygen isotope effect on $T_c$}

The oxygen isotope effect on $T_c$ is determined by the $\alpha$ isotope exponent which is defined as:

\begin{eqnarray}
  \alpha\para{x}=-\frac{d \log T_{c}\para{x}}{d\log M_{_O}}
\end{eqnarray}

In the underdoped region, phase fluctuation studies show that $T_c$ is controlled by the superfluid density {\em i.e.} $T_c\sim T_{\rm BKT} \propto \rho_s \propto 1/m^*_h$. Therefore, we have:

\begin{eqnarray}
  &&x<x_{\rm OD}: \cr
  &&\alpha\para{x}=\frac{d \log m^*_h\para{x,T_{c}}}{d\log M_{_{O}}}=q^2_{h}\para{x,T_{c}}\frac{\lambda_{0}}{\omega_{_{E}}},\sd
\end{eqnarray}
where $x_{\rm OD}$ stands for the optimal doping. Since $q_{h}\para{x,T_{c}}\propto \rho_{f}\para{T_{c}}$ is a decreasing function of doping and at the optimal doping and beyond vanishes, isotope exponent dies off as we approach the optimal doping and finally becomes negligible. On the overdoped cuprates, at the boundary between the SC and FL phases $q_h$ vanishes and therefore, we do not expect isotope effect on $T_c$ for that case. These conclusion are also consistent with experimental observations.

\section{Summary and conclusion}
We have studied the electron-phonon interaction in the strong correlation regime within the spin charge separation framework. It was shown that by the careful study of
the gauge theory and by taking local constraints into account, the electric charge of slave particles can be determined unambiguously. The charge of holons was calculated and shown to be proportional to the superfluid density of spinons in the SC region. Furthermore, the charge vanishes in the FL region whereas it grows in the PG phase. We argued that holons couple to phonons due to their nonzero electric charge. The holon-phonon interaction should be treated in the non-adiabatic limit since holons are bosonic quasiparticles and their energy is much less than that of phonons as long as they condense. The implications of our arguments on the OIE in high Tc cuprates were discussed above, and we have shown that the effective charge of holons and the exponential dependence of their mass on $q_h$ readily explains the unusual  $^{16}$O/$^{18}$ isotope effect in cuprates. Our model successfully explains the general trend in experiments and supports the spin charge separation idea.

%

\end{document}